\documentclass[aps,prb,onecolumn,floatfix,superscriptaddress]{revtex4}
\usepackage{graphicx}
\usepackage[english]{babel}
\usepackage{amsmath}
\usepackage{amssymb}
\usepackage{slashed}
\usepackage{bm}

\newcommand{\me}{\mathrm{e}}

\newcommand{\dif}{\mathrm{d}}

\newcommand{\bk}{\mathbf{k}}
\newcommand{\bp}{\mathbf{p}}
\newcommand{\bq}{\mathbf{q}}

\newcommand{\bl}{\mathbf{l}}
\allowdisplaybreaks
\begin{document}

\title{BCS thermal vacuum of fermionic superfluids and its perturbation theory }

\author{Xu-Yang Hou}
\affiliation{Department of Physics, Southeast University, Jiulonghu Campus, Nanjing 211189, China}
\author{Ziwen Huang}
\affiliation{Department of Physics, Southeast University, Jiulonghu Campus, Nanjing 211189, China}
\author{Hao Guo}
\email{guohao.ph@seu.edu.cn}
\affiliation{Department of Physics, Southeast University, Jiulonghu Campus, Nanjing 211189, China}
\author{Yan He}
\affiliation{College of Physical Science and Technology, Sichuan University, Chengdu, Sichuan 610064, China}
\author{Chih-Chun Chien}
\email{cchien5@ucmerced.edu}
\affiliation{School of Natural Sciences, University of California, Merced, CA 95343, USA}

\begin{abstract}
The thermal field theory is applied to fermionic superfluids by doubling the degrees of freedom of the BCS theory. We construct the two-mode states and the corresponding Bogoliubov transformation to obtain the BCS thermal vacuum. The expectation values with respect to the BCS thermal vacuum produce the statistical average of the thermodynamic quantities. The BCS thermal vacuum allows a quantum-mechanical perturbation theory with the BCS theory serving as the unperturbed state. We evaluate the leading-order corrections to the order parameter and other physical quantities from the perturbation theory. A direct evaluation of the pairing correlation as a function of temperature shows the pseudogap phenomenon results from the perturbation theory. The BCS thermal vacuum is shown to be a generalized coherent and squeezed state. The correspondence between the thermal vacuum and purification of the density matrix allows a unitary transformation, and we found the geometric phase in the parameter space associated with the transformation.
\end{abstract}

\maketitle

\section{Introduction}
Quantum many-body systems can be described by quantum field theories~\cite{AGD_book,Fetter_book,Kapusta_book,Wen_book}. Some available frameworks for systems at finite temperatures include the Matsubara formalism using the imaginary time for equilibrium systems~\cite{Matsubara,AGD_book} and the Keldysh formalism of time-contour path integrals~\cite{Keldysh,Kapusta_book} for non-equilibrium systems. There are also alternative formalisms. For instance, the thermal field theory~\cite{TFD_book,TFD_book2,Vitiello_book,Das_book} is built on the concept of thermal vacuum.  The idea of thermal vacuum is to construct temperature-dependent augmented states and rewrite the statistical average of observables as quantum-mechanical expectation values. Thermal field theory was introduced a while ago~\cite{TFD_book,Umezawa_book}, and more recently it has found applications beyond high-energy physics~\cite{Vitiello_book}.

The thermal vacuum of a non-interacting bosonic or fermionic system has been obtained by a Bogoliubov transformation of the corresponding two-mode vacuum~\cite{Unruh_book}, where an auxiliary system, called the tilde system, is introduced to satisfy the statistical weight. The concepts of Bogoliubov transformation and unitary inequivalent representations are closely related to the development of thermal vacuum theory~\cite{Vitiello_book}, and they are also related to the quantum Hall effect~\cite{Ezawa_book} and orthogonality catastrophe~\cite{Wen_book}. The thermal vacuum of an interacting system can be constructed if a Bogoliubov transformation of the corresponding two-mode vacuum is found. In the following, we will use the BCS theory of fermionic superfluids as a concrete example. By construction, the thermal vacuum provides an alternative interpretation of the statistical average. Nevertheless, we will show that the introduction of the thermal vacuum simplifies certain calculations from the level of quantum field theory to the level of quantum mechanics. As an example, we will apply the thermal field theory to develop a perturbation theory where the BCS thermal vacuum serves as the unperturbed state, and the fermion-fermion interaction ignored in the BCS approximation is the perturbation. Because the particle-hole channel, or the induced interaction~\cite{Heiselberg00}, is not included in the BCS theory, the BCS thermal vacuum inherits the same property and its perturbation theory does not produce the Gorkov-Melik-Barkhudarov effect~\cite{GMB61} predicting the transition temperature is more than halved.

The thermal vacuum of a given Hamiltonian can alternatively be viewed as a purification of the density matrix of the corresponding mixed state~\cite{IssacChuang_book,GPhase_book}. Since the thermal vacuum is a pure state, the statistical average of a physical quantity at finite temperatures is the expectation value obtained in the quantum mechanical manner. Therefore, one can find some physical quantities not easily evaluated in conventional methods. This is the reason behind the perturbation theory using the BCS thermal vacuum as the unperturbed state. The corrections from the full fermion-fermion interaction can be evaluated order-by-order following the standard time-independent quantum-mechanical perturbation theory~\cite{MQM}. In principle, the corrections to any physical quantities can be expressed as a perturbation series. In this work, the first-order corrections to the order parameter will be evaluated and analyzed.

Furthermore, the BCS thermal vacuum allows us to gain insights into the BCS superfluids. We will illustrate the applications of the BCS thermal vacuum by calculating the pairing correlation, which may be viewed as the quantum correlation matrix~\cite{Qi18} of the pairing operator. There is interest in finding the parent Hamiltonian of trial states~\cite{Qi18,Greiter18}. We found that the corrected order parameter vanishes at a lower temperature compared to the unperturbed one, and the pairing correlation persists above the corrected transition temperature. Thus, the perturbation theory offers an algebraic foundation for the pseudogap effect~\cite{Ourreview,OurAnnPhys}, where the pairing effect persists above the transition temperature.

The quantum-mechanical level calculations using the thermal BCS vacuum show that the BCS thermal vacuum saturates a generalized Heisenberg uncertainty relation and is a generalized squeezed coherent state. Moreover, the purification of the density matrix of a mixed state allows a unitary transformation~\cite{IssacChuang_book,GPhase_book}. For the BCS thermal vacuum, this translates to a $U(1)$ phase in its construction. By evaluating the analogue of the Berry phase~\cite{Berry84} along the $U(1)$-manifold of the unitary transformation, we found a thermal phase characterizing the thermal excitations of the BCS superfluid.

The rest of the paper is organized as follows. In Sec.~\ref{InTFD}, we briefly review the general formalism of thermal field theory. Sec.~\ref{TVBCS} shows the construction of the BCS thermal vacuum from the BCS theory and some properties of the BCS thermal vacuum. In Sec.~\ref{PABCS}, we present the perturbation theory using the BCS thermal vacuum as the unperturbed state and derive the first-order corrections to thermodynamic quantities and the BCS order parameter.
Sec.~\ref{secV} presents three applications of the BCS thermal vacuum. We evaluate the pairing correlation and show that it persists above the superfluid transition temperature. Thus, the pseudogap effect arises in the perturbation theory. We also show that the BCS thermal vacuum saturates a generalized Heisenberg inequality, and there is a thermal phase associated with the unitary transformation of the BCS thermal vacuum.
The conclusion is given in Sec~\ref{conclusion}. The details of our calculations are summarized in the Appendix.

\section{A Brief Review of Thermal Field Theory}\label{InTFD}
Throughout this paper, we choose $\hbar=k_B=1$ and set $|q_e|=1$, where $q_e$ is the electron charge.
In quantum statistics, the expectation value of an operator $\mathcal{A}$ in a canonical ensemble is evaluated by the statistical (thermal) average
\begin{equation}\label{TFD0}
\langle \mathcal{A}\rangle_\beta\equiv \frac{1}{Z}\mathrm{Tr} (\mathrm{e}^{-\beta H}\mathcal{A}),
\end{equation}
where $Z=\mathrm{Tr}\mathrm{e}^{-\beta H}$ is the partition function with the Hamiltonian $H$ at temperature $T\equiv\frac{1}{\beta}$.
A connection between the statistical average and quantum field theory is established by rewriting the partition function in the path-integral form and introducing the imaginary time $\tau=it$. This method allows a correspondence between the partition functions in statistical mechanics and quantum field theory~\cite{Wen_book,Kapusta_book}. Therefore, one usually focuses on equilibrium statistical mechanics~\cite{Fetter_book}, where the Matsubara frequencies are introduced~\cite{Matsubara}.

The central idea of thermal field theory is to
express the statistical average
over a set of mixed quantum states as the expectation value of a temperature-dependent pure state, called the thermal vacuum $|0(\beta)\rangle$~\cite{QP_book,Vitiello_book}. Explicitly,
\begin{equation}\label{TFD1}
\langle \mathcal{A}\rangle_\beta=\langle0(\beta)|\mathcal{A}|0(\beta)\rangle=Z^{-1}\sum_n\langle n|\mathcal{A}|n\rangle \mathrm{e}^{-\beta E_n}.
\end{equation}
Here $|n\rangle$ are the orthonormal eigenstates of the Hamiltonian $H$ satisfying $\langle n|m\rangle=\delta_{nm}$ with eigenvalues $E_n$.
Following Refs.~\cite{QP_book,TFD_book}, one way to achieve this is to double the degrees of freedom of the system by introducing an auxiliary system identical to the one we are studying. This auxiliary system is usually denoted by the tilde symbol, so its Hamiltonian is $\tilde{H}$ and its eigenstates $|\tilde{n}\rangle$ satisfy $\tilde{H}|\tilde{n}\rangle=E_n|\tilde{n}\rangle$ and $\langle\tilde{n}|\tilde{m}\rangle=\delta_{nm}$.

It is imposed that th non-tilde bosonic (fermionic) operators commute (anti-commute) with their tilde counterparts~\cite{QP_book,TFD_book}. To find $\vert 0(\beta)\rangle$, one needs to consider the space spanned by the direct product of the tilde and non-tilde state-vectors $\vert n,\tilde{m}\rangle\equiv\vert n\rangle\otimes\vert \tilde{m}\rangle$. Hence, the thermal vacuum can be expressed as
\begin{equation}\label{tv1}
\vert 0(\beta)\rangle=\sum_n\vert n\rangle \otimes|f_n(\beta)\rangle= \sum_n f_n(\beta)\vert n,\tilde{n}\rangle,
\end{equation}
where $|f_n(\beta)\rangle=f_n(\beta)|\tilde{n}\rangle$ with $f_n^*(\beta)f_m(\beta)=Z^{-1}\mathrm{e}^{-\beta E_n}\delta_{nm}$.
It is also required that the operator $\mathcal{A}$, whose expectation values we are interested in, only acts on the non-tilde space~\cite{QP_book,TFD_book}. Then, $\langle 0(\beta)\vert A\vert 0(\beta)\rangle=\sum\limits_n |f_n(\beta)|^2\langle n\vert \mathcal{A}\vert n\rangle.$
Eq.~(\ref{TFD1}) is satisfied if
\begin{eqnarray}\label{t80}
|f_n(\beta)|^2=Z^{-1}\mathrm{e}^{-\beta E_n}.
\end{eqnarray}
Therefore, the thermal vacuum can be expressed as
\begin{equation}\label{GDM}
\vert 0(\beta)\rangle=\sum_n \frac{1}{\sqrt{Z}}\mathrm{e}^{-\beta\frac{E_n}{2}+i\chi_n}\vert n,\tilde{n}\rangle,
\end{equation}
where we introduce an unitary factor $\exp(i\chi_n)$ to each coefficient $f_n(\beta)$.  This is allowed because one may view the thermal vacuum as a purification of the density matrix, and different purified states of the same density matrix can be connected by a unitary transformation~\cite{IssacChuang_book,GPhase_book}.

The thermal vacuum is usually not the ground state of either $H$ or $\tilde{H}$, but it is the zero-energy eigenstate of $\hat{H}\equiv H-\tilde{H}$. Since the energy spectrum of $\hat{H}$ has no lower bound, it is not physical. Nevertheless, if the thermal vacuum can be constructed by performing a unitary transformation $U(\beta)$ on the two-mode ground state, $\vert 0(\beta)\rangle=U(\beta)\vert 0,\tilde{0}\rangle$,
then the thermal vacuum is an eigenstate of the ``thermal Hamiltonian'' $H(T)\equiv U(\beta)HU^{-1}(\beta)$ with eigenvalue $E_0$, which is also the ground-state energy of $H$. This is because
$H(T)\vert 0(\beta)\rangle=UH\vert 0,\tilde{0}\rangle=E_0U\vert 0,\tilde{0}\rangle=E_0\vert 0(\beta)\rangle$.
We remark that $U(\beta)$ may contain creation and annihilation operators.
Importantly, $\vert 0(\beta)\rangle$ is also the ground state of $H(T)$ since unitary transformations do not change the eigenvalues of an operator. The origin of the name ``thermal vacuum'' comes from the fact that it is the finite-temperature generalization of the two-mode ground state\cite{TFD_book,TFD_book2}. 

\section{BCS Thermal Vacuum}\label{TVBCS}
We first give a brief review of the BCS theory.
The fermion field operators $\psi_\sigma$ and $\psi^{\dagger}_\sigma$ satisfy
\begin{align}
&\{\psi_{\bk\sigma},\psi^\dagger_{\bk'\sigma'}\}=\delta_{\bk\bk'}\delta_{\sigma\sigma'}, \quad \sigma,\sigma'=\uparrow,\downarrow
\end{align}
and all other anti-commutators vanish.
The Hamiltonian of a two-component Fermi gas with attractive contact interactions is given by
\begin{eqnarray}\label{Htotal}
H=\sum_{\bk}\psi^\dag_{\bk\sigma}(\frac{\bk^2}{2m}-\mu)\psi_{\bk\sigma}-g\sum_{\bk\bp\bq}\psi^\dag_{\bk-\bp\uparrow}\psi^\dag_{\bp\downarrow}\psi_{\bk-\bq\downarrow}\psi_{\bq\uparrow},
\end{eqnarray}
where $m$ and $\mu$ are the fermion mass and chemical potential, respectively, $g$ is the coupling constant, and $\sum_{\bk}=\mathcal{V}\int\frac{\dif^3\bk}{(2\pi)^3}$ with $\mathcal{V}$ being the volume of the system.
In the BCS theory, the pairing field leads to the gap function  $\Delta(\mathbf{x})=g\langle\psi_{\uparrow}(\mathbf{x})\psi_{\downarrow}(\mathbf{x})\rangle$, which is also the order parameter in the broken-symmetry phase. Physically, pairing between fermions in the time-reversal states $(\bk\uparrow)$ and ($-\bk\downarrow$) can make the Fermi sea unstable if the inter-particle interaction is attractive~\cite{Tinkham_SCbook}.

The Hamiltonian is then approximated by the BCS form~\cite{Schrieffer_book,Walecka}
\begin{align}\label{HBCS}
H_{\textrm{BCS}}=\frac{|\Delta|^2}{g}+\sum_{\mathbf{k}\sigma}\psi^{\dagger}_{\mathbf{k}\sigma}\Big(\frac{\mathbf{k}^2}{2m}-\mu\Big)\psi_{\mathbf{k}\sigma}
+\sum_{\mathbf{k}}(\Delta^*\psi_{-\mathbf{k}\uparrow}\psi_{\mathbf{k}\downarrow}
+\Delta\psi^{\dagger}_{\mathbf{k}\downarrow}\psi^{\dagger}_{-\mathbf{k}\uparrow}),
\end{align}
which can be diagonalized as
\begin{align}\label{HBCS2}
H_{\textrm{BCS}}=\sum_{\mathbf{k}}E_{\mathbf{k}}(\alpha^{\dagger}_{\mathbf{k}}\alpha_{\mathbf{k}}+\beta^{\dagger}_{-\mathbf{k}}\beta_{-\mathbf{k}})+\sum_{\mathbf{k}}(\xi_{\mathbf{k}}-E_{\mathbf{k}})+\frac{|\Delta|^2}{g}.
\end{align}
This is achieved by implementing the Bogoliubov transformation, which can also be cast in the form of a similarity transformation:
\begin{align}\label{BT2}
\alpha_{\mathbf{k}}=\me^G \psi_{\bk\uparrow}\me^{-G}
=\cos\phi_{\mathbf{k}}\psi_{\mathbf{k}\uparrow}-\sin\phi_{\mathbf{k}}\psi^{\dagger}_{-\mathbf{k}\downarrow},\quad
\beta^{\dagger}_{-\mathbf{k}}=\me^G \psi_{-\bk\downarrow}\me^{-G}
=\sin\phi_{\mathbf{k}}\psi_{\mathbf{k}\uparrow}+\cos\phi_{\mathbf{k}}\psi^{\dagger}_{-\mathbf{k}\downarrow}.
\end{align}
Here $\cos(2\phi_{\mathbf{k}})=\frac{\xi_{\mathbf{k}}}{E_{\mathbf{k}}}$, $\sin(2\phi_{\mathbf{k}})=\frac{|\Delta|}{E_{\mathbf{k}}}$, $E_{\mathbf{k}}=\sqrt{\xi_{\mathbf{k}}^2+|\Delta|^2}$ is the quasi-particle dispersion, $G=\sum_\bk\phi_\bk(\psi_{\bk\uparrow}\psi_{-\bk\downarrow}+\psi^\dagger_{\bk\uparrow}\psi^\dagger_{-\bk\downarrow})$ is the generator of the transformation. Since $G^\dagger=-G$, $e^{-G}$ is unitary. The field operators of the quasi-particles satisfy the anti-commutation relations
\begin{align}
&\{\alpha_\bk,\alpha^\dagger_{\bk'}\}=\delta_{\bk\bk'},\{\beta_\bk,\beta^\dagger_{\bk'}\}=\delta_{\bk\bk'},
\end{align}
and all other anti-commutators vanish. We remark that the BCS theory only considers the particle-particle channel (pairing) contribution. As pointed out in Ref.~\cite{Mihaila11}, the BCS theory is not compatible with a split, density contribution to the chemical potential. It is, however, possible to add the particle-hole (density) diagrams to the Feynman diagrams describing the scattering process~\cite{Heiselberg00} and obtain a modified effective interaction, which then leads to the Gorkov-Melik-Barkhudarov effect~\cite{GMB61} of suppressed superfluid transition temperature. Here we base the theory on the BCS theory, so the resulting thermal field theory also does not exhibit the particle-hole channel effects.

Rewriting the Bogoliubov transformation as a similarity transformation leads to a connection between the Fock-space vacuum $|0\rangle$ of the $\psi_{\sigma}$ quanta and the BCS ground state $|g\rangle$, which can be viewed as the vacuum of the $\alpha_{\mathbf{k}}$ and $\beta_{-\mathbf{k}}$ quasi-particles because $\alpha_{\mathbf{k}}|g\rangle=0=\beta_{-\mathbf{k}}|g\rangle$. Explicitly,
\begin{align}
\vert g\rangle =\me^G|0\rangle
=\prod\limits_{\mathbf{k}}(\cos\phi_{\mathbf{k}}+\sin\phi_{\mathbf{k}}\psi^\dagger_{\mathbf{k}\uparrow}\psi^\dagger_{-\mathbf{k}\downarrow})\vert 0\rangle.
\end{align}
We remark the relation between the similarity transformation of the fields and the unitary transformation of the Fock-space vacuum resembles the connection between the Schrodinger picture and the Heisenberg picture in quantum dynamics (see Ref.~\onlinecite{MQM} for example).

\subsection{Constructing BCS thermal vacuum}
The thermal vacuum of the BCS theory is constructed by introducing the tilde partners of the $\alpha_{\mathbf{k}}$ and $\beta_{-\mathbf{k}}$ quanta, $\tilde{\alpha}_{\mathbf{k}}$ and $\tilde{\beta}_{-\mathbf{k}}$. They satisfy the algebra
\begin{align}
\{\tilde{\alpha}_\bk,\tilde{\alpha}^\dagger_{\bk'}\}=\delta_{\bk\bk'},\{\tilde{\beta}_\bk,\tilde{\beta}^\dagger_{\bk'}\}=\delta_{\bk\bk'},
\end{align}
and all other anti-commutators vanish. Moreover, the tilde fields anti-commute with the quasi-particle quanta $\alpha_\bk$ and $\beta_\bk$.
Next, the two-mode BCS ground state is constructed as follows.
\begin{align}
|g,\tilde{g}\rangle
=\prod_{\mathbf{k}}\vert 0,\tilde{0}\rangle_{\alpha_{\mathbf{k}}}\otimes\vert 0,\tilde{0}\rangle_{\beta_{-\mathbf{k}}}.
\end{align}
Here $\vert 0,\tilde{0}\rangle_{\alpha_{\mathbf{k}}}$ ($\vert 0,\tilde{0}\rangle_{\beta_{-\mathbf{k}}}$) denotes the Fock-space vacuum of $\alpha_{\mathbf{k}}$ and $\tilde{\alpha}_{\mathbf{k}}$ ($\beta_{-\mathbf{k}}$ and $\tilde{\beta}_{-\mathbf{k}}$).

The occupation number of each fermion quasi-particle state can only be $0$ or $1$.  According to Eq.~(\ref{tv1}), the two-mode BCS thermal vacuum can be expressed as
\begin{align}
\label{ct212_1}
\vert 0(\beta)\rangle=\prod_{\mathbf{k}}\otimes (f_{0\mathbf{k}}\vert 0,\tilde{0}\rangle_{\alpha_{\mathbf{k}}}+f_{1\mathbf{k}}\vert 1,\tilde{1}\rangle_{\alpha_{\mathbf{k}}})\otimes (f_{0\mathbf{k}}\vert 0,\tilde{0}\rangle_{\beta_{-\mathbf{k}}}+f_{1\mathbf{k}}\vert 1,\tilde{1}\rangle_{\beta_{-\mathbf{k}}}),
\end{align}
where $\vert 1,\tilde{1}\rangle_{\alpha_{\mathbf{k}}}=\alpha^{\dagger}_{\mathbf{k}}\tilde{\alpha}^{\dagger}_{\mathbf{k}}|g,\tilde{g}\rangle$ and $\vert 1,\tilde{1}\rangle_{\beta_{-\mathbf{k}}}=\beta^{\dagger}_{-\mathbf{k}}\tilde{\beta}^{\dagger}_{-\mathbf{k}}|g,\tilde{g}\rangle$.
The coefficients $f_{0\mathbf{k}}$ and $f_{1\mathbf{k}}$ can be deduced from Eq.~(\ref{t80}).
For each $\mathbf{k}$, we define $Z_\mathbf{k}=1+\mathrm{e}^{-\beta E_\mathbf{k}}$ and then the partition function is  $Z=\prod_\mathbf{k}Z_\mathbf{k}$. Comparing with Eq.~(\ref{t80}), we get
\begin{align}\label{eq:f0f1}
|f_{0\mathbf{k}}|=Z_\mathbf{k}^{-\frac{1}{2}}=\frac{1}{\sqrt{1+\mathrm{e}^{-\beta E_\mathbf{k}}}},\quad |f_{1\mathbf{k}}|=Z_\mathbf{k}^{-\frac{1}{2}}\mathrm{e}^{-\beta E_\mathbf{k}/2}=\frac{1}{\sqrt{1+\mathrm{e}^{\beta E_\mathbf{k}}}}.
\end{align}
According to Eq.~(\ref{GDM}), one may choose a relative phase between the different two-mode states. Here, we choose $\chi_0=0$ and $\chi_1=-\chi$. Thus,
the coefficients are parametrized by
\begin{equation}\label{f0f1}
f_{0\mathbf{k}}=\cos\theta_{\mathbf{k}},~ f_{1\mathbf{k}}=\sin\theta_{\mathbf{k}}\mathrm{e}^{-i\chi}.
\end{equation}
The phases $\chi$ parametrizes the U(1) transformation allowed by the BCS thermal vacuum, and we will show its consequence later.

The BCS thermal vacuum can be obtained by a unitary transformation of the two-mode BCS ground state. Explicitly,
\begin{align}
\label{ct212_3}
\vert 0(\beta)\rangle&=\mathrm{e}^Q\vert g,\tilde{g}\rangle
=\prod\limits_{\mathbf{k}}(\cos\theta_{\mathbf{k}}+\sin\theta_{\mathbf{k}}\mathrm{e}^{-i\chi}\alpha_{\mathbf{k}}^\dagger\tilde{\alpha}_{\mathbf{k}}^\dagger)(\cos\theta_{\mathbf{k}}+\sin\theta_{\mathbf{k}}\mathrm{e}^{-i\chi}\beta_{-\mathbf{k}}^\dagger\tilde{\beta}_{-\mathbf{k}}^\dagger)\vert g,\tilde{g}\rangle,
\end{align}
where
\begin{align}
\label{ct212_2}
Q=\sum_\mathbf{k}\theta_\mathbf{k}(\alpha_{\mathbf{k}}\tilde{\alpha}_{\mathbf{k}}\mathrm{e}^{i\chi}+\alpha_{\mathbf{k}}^\dagger \tilde{\alpha}_{\mathbf{k}}^\dagger \mathrm{e}^{-i\chi}+\beta_{-\mathbf{k}}\tilde{\beta}_{-\mathbf{k}}\mathrm{e}^{i\chi}+\beta_{-\mathbf{k}}^\dagger \tilde{\beta}_{-\textbf{k}}^\dagger \mathrm{e}^{-i\chi}).
\end{align}
Following a similarity transformation using the unitary operator $\mathrm{e}^{-Q}$, the BCS thermal vacuum is the Fock-space vacuum of the thermal quasi-quanta
\begin{eqnarray}\label{eq:BogoT}
\alpha_{\mathbf{k}}(T)=\mathrm{e}^{Q}\alpha_{\mathbf{k}}\mathrm{e}^{-Q}=\alpha_{\mathbf{k}} \cos\theta_{\mathbf{k}}-\tilde{\alpha}_{\mathbf{k}}^\dagger\sin\theta_{\mathbf{k}}\mathrm{e}^{-i\chi},\quad
\beta_{-\mathbf{k}}(T)=\mathrm{e}^{Q}\beta_{\mathbf{k}}\mathrm{e}^{-Q}=\beta_{-\mathbf{k}} \cos\theta_{\mathbf{k}}-\tilde{\beta}_{\mathbf{k}}^\dagger\sin\theta_{\mathbf{k}}\mathrm{e}^{-i\chi}.
\end{eqnarray}
This is because $\alpha_{\mathrm{k}}(T)\vert 0(\beta)\rangle=\mathrm{e}^{Q}\alpha_{\mathrm{k}}\vert g,\tilde{g}\rangle=0$ and $\beta_{-\mathrm{k}}(T)\vert 0(\beta)\rangle=\mathrm{e}^{Q}\beta_{-\mathrm{k}}\vert g,\tilde{g}\rangle=0$. One can construct similar relations for the tilde operators. Moreover, $|0(\beta)\rangle$ is the ground state of the thermal BCS Hamiltonian
\begin{align}\label{HBCST}
H_{\textrm{BCS}}(T)\equiv \textrm{e}^QH_{\textrm{BCS}}\textrm{e}^{-Q}=\sum_{\mathbf{k}}E_{\mathbf{k}}\big[\alpha^{\dagger}_{\mathbf{k}}(T)\alpha_{\mathbf{k}}(T)+\beta^{\dagger}_{-\mathbf{k}}(T)\beta_{-\mathbf{k}}(T)\big]+\sum_{\mathbf{k}}(\xi_{\mathbf{k}}-E_{\mathbf{k}})+\frac{|\Delta|^2}{g}.
\end{align}
Incidentally, the similarity transformation does not change the eigenvalues.

At zero temperature, $|f_{0\mathbf{k}}|=1$ and $f_{1\mathbf{k}}=0$. Hence, the BCS thermal vacuum reduces to the ground state of the conventional BCS theory, but in the augmented two-mode form. It is important to notice that the two-mode BCS ground state differs from the BCS thermal vacuum at finite temperatures in their structures: The former is the Fock-space vacuum of the quasi-particles, i.e., $\alpha_\mathbf{k}|g,\tilde{g}\rangle=\beta_\mathbf{k}|g,\tilde{g}\rangle=0$; the latter is the Fock-space vacuum of the thermal quasi-particles, i.e., $\alpha_\mathbf{k}(T)|0(\beta)\rangle=\beta_\mathbf{k}(T)|0(\beta)\rangle=0$.

\subsection{Equations of State}
By construction, the statistical average of any physical observable can be obtained by taking the expectation value with respect to the thermal vacuum. In the following we show how this procedure reproduces the BCS number and gap equations. One can show that
\begin{eqnarray}\label{E3}
\langle 0(\beta)| \alpha_{\mathbf{k}}^\dagger\alpha_{\mathbf{k}} |0(\beta)\rangle
&=&\langle g,\tilde{g}\vert \mathrm{e}^{-Q}\alpha_{\mathbf{k}}^\dagger\alpha_{\mathbf{k}}\mathrm{e}^Q\vert g,\tilde{g}\rangle
=f(E_{\mathbf{k}}).
\end{eqnarray}
Here we have used Eq.~\eqref{eq:BogoT} and $\sin^2\theta_\mathbf{k}=f(E_{\mathbf{k}})$.
Similarly, one can show that
$\langle0(\beta)| \beta_{-\mathbf{k}}^\dagger\beta_{-\mathbf{k}}|0(\beta)\rangle=f(E_{\mathbf{k}})$, $\langle0(\beta)| \beta_{\mathbf{-k}}\beta_{\mathbf{-k}}^\dagger|0(\beta)\rangle=\langle0(\beta)| \alpha_{\mathbf{k}}\alpha^\dagger_{\mathbf{k}}|0(\beta)\rangle=1-f(E_{\mathbf{k}})$, and $\langle0(\beta)| \beta_{\mathbf{-k}}\alpha_{\mathbf{k}}|0(\beta)\rangle=\langle0(\beta)|\alpha^\dagger_{\mathbf{k}}\beta^\dagger_{\mathbf{-k}}|0(\beta)\rangle=0$.
Applying these identities and the inverse transformation of Eq.~(\ref{BT2}),
the total particle number is given by the expectation value of the number operator with respect to the state $\vert 0(\beta)\rangle$:
\begin{align}\label{Neqn}
N&=\sum_\mathbf{k} \big(\langle0(\beta)|\psi_{\mathbf{k}\uparrow}^\dagger\psi_{\mathbf{k}\uparrow}|0(\beta)\rangle+\langle0(\beta)| \psi_{-\mathbf{k}\downarrow}^\dagger\psi_{-\mathbf{k}\downarrow}|0(\beta)\rangle\big)\nonumber\\
&=\sum_{\mathbf{k}}\big[\langle0(\beta)|(|u_{\mathbf{k}}|^2\alpha_{\mathbf{k}}^\dagger\alpha_{\mathbf{k}}+|v_{\mathbf{k}}|^2\beta_{\mathbf{-k}}\beta^\dagger_{\mathbf{-k}}+v_{\mathbf{k}}u_{\mathbf{k}}\beta_{\mathbf{-k}}\alpha_{\mathbf{k}}+u_{\mathbf{k}}^*v_{\mathbf{k}}^*\alpha_{\mathbf{k}}^\dagger\beta_{\mathbf{-k}}^\dagger)|0(\beta)\rangle\nonumber\\
& \quad+\langle0(\beta)|(|u_{\mathbf{k}}|^2\beta_{\mathbf{-k}}^\dagger\beta_{\mathbf{-k}}+|v_{\mathbf{k}}|^2\alpha_{\mathbf{k}}\alpha^\dagger_{\mathbf{k}}-v_{\mathbf{k}}u_{\mathbf{k}}\alpha_{\mathbf{k}}\beta_{\mathbf{-k}}-u_{\mathbf{k}}^*v_{\mathbf{k}}^*\beta_{\mathbf{-k}}^\dagger\alpha_{\mathbf{k}}^\dagger)|0(\beta)\rangle\big]\notag\\
&=2\sum_{\mathbf{k}}\Big\{|u_{\mathbf{k}}|^2f(E_{\mathbf{k}})+|v_{\mathbf{k}}|^2\big[1-f(E_{\mathbf{k}})\big]\Big\}.
\end{align}
Here $|u_\bk|^2,|v_\bk|^2=(1\pm\xi_\bk/E_\bk)/2$, and the expression is the same as the one from the finite-temperature Green's function formalism~\cite{Fetter_book}. The gap equation can be deduced in a similar way. Explicitly,
\begin{align}\label{Geqn}
\Delta=-g\sum_\mathbf{k}\langle0(\beta)| \psi_{\mathbf{k}\uparrow}\psi_{\mathbf{-k}\downarrow}|0(\beta)\rangle=g\Delta\sum_{\mathbf{k}}\frac{1-2f(E_{\mathbf{k}})}{2E_{\mathbf{k}}}.
\end{align}
When compared to the Green's function approach~\cite{Fetter_book}, the thermal-vacuum approach is formally at the quantum mechanical level. To emphasize this feature, we will use some techniques from quantum mechanics to perform calculations equivalent to their complicated counterparts in the framework of field field theory.

\section{Perturbation Theory Based on BCS Thermal Vacuum}\label{PABCS}
The BCS equations of state can be derived from a formalism formally identical to quantum mechanics by the BCS thermal vacuum. We generalize the procedure to more complicated calculations such as evaluating higher-order corrections to the BCS mean-field theory by developing a perturbation theory like the one in quantum mechanics. The idea is to take the BCS thermal vacuum as the unperturbed state and follow the standard time-independent perturbation formalism~\cite{MQM} to build the corrections order by order.

\subsection{Basic Framework}
To develop a perturbation theory at the quantum mechanical level based on the BCS thermal vacuum, we first identify the omitted interaction term in the BCS approximation.
By comparing the total Hamiltonian (\ref{Htotal}) and the BCS Hamiltonian (\ref{HBCS}), one finds
\begin{equation}\label{5}
H=H_\textrm{BCS}-g\sum_{\bk\neq 0}\sum_{\bp\bq}\psi^\dag_{\bk-\bp\uparrow}\psi^\dag_{\bp\downarrow}\psi_{\bk-\bq\downarrow}\psi_{\bq\uparrow},
\end{equation}
where the second term can be thought of as a perturbation to the BCS Hamiltonian. Therefore, we take $H_0\equiv H_\textrm{BCS}$ as the unperturbed Hamiltonian since the BCS ground state is known. The perturbation is
\begin{equation}\label{pV}
V=-g\sum_{\bk\neq 0}\sum_{\bp\bq}\psi^\dag_{\bk-\bp\uparrow}\psi^\dag_{\bp\downarrow}\psi_{\bk-\bq\downarrow}\psi_{\bq\uparrow}.
\end{equation}

Next, the BCS thermal vacuum is used to find the contributions from the perturbation. The BCS thermal vacuum $|0(\beta)\rangle$ is the ground state of the unperturbed thermal BCS Hamiltonian $H_\textrm{BCS}(T)=\me^QH_\textrm{BCS}\me^{-Q}$. Our task is to find the thermal vacuum $|0(\beta)\rangle_\textrm{c}$ of the total thermal Hamiltonian
\begin{equation}
H(T)\equiv H_\textrm{BCS}(T)+V(T)=\me^QH_\textrm{BCS}\me^{-Q}+\me^QV\me^{-Q}.
\end{equation}
Following the perturbation theory in quantum mechanics, the full thermal vacuum has the structure
\begin{align}\label{CTVC}
|0(\beta)\rangle_\textrm{c}
&=|0(\beta)\rangle+\sum_{k\neq 0}|k^{(0)}\rangle\frac{V_{k0}}{E^{(0)}_0-E^{(0)}_k}+\cdots,
\end{align}
where $ V_{k0}=\langle k^{(0)}|V|0(\beta)\rangle $ is the matrix element of the perturbation with respect to the unperturbed states, $|k^{(0)}\rangle$ includes all possible unperturbed excited states given by $\alpha^\dag_\bk(T)|0(\beta)\rangle$, $\beta^\dag_{-\bk}(T)|0(\beta)\rangle$, $\beta^\dag_{-\bk}(T)\alpha^\dag_\bk(T)|0(\beta)\rangle$, etc., and $E^{(0)}_n$ is the corresponding unperturbed energy. After obtaining the full BCS
thermal vacuum order by order, the corrections to physical quantities such as the order parameter can be found by taking the expectation values of the corresponding operators with respect to the perturbed thermal vacuum.

\subsection{Corrections to physical quantities}
The perturbation theory requires the evaluation of the matrix elements of the perturbation, $V_{k0}$, $k=1,2,\cdots$, which are determined as follows. Let $|k^{(0)}\rangle=\mathcal{O}_k^{\dagger}(T)\cdots\mathcal{O}_1^{\dagger}(T)|0(\beta)\rangle$ be a $k$-particle excited state, where $\mathcal{O}_i$ represents the quasi-particle annihilation operator $\alpha_{\bk_i}$ or $\beta_{-\bk_i}$.
Then,
\begin{align}\label{Vk0}
V_{k0}&=\langle k^{(0)}|V(T)|0(\beta)\rangle\notag \\
&=\langle g,\tilde{g} |\me^{-Q}\mathcal{O}_1(T)\me^Q\me^{-Q}\cdots\me^{Q}\me^{-Q}\mathcal{O}_n(T)\me^{Q}\me^{-Q}\me^QV\me^{-Q}\me^Q|g,\tilde{g}\rangle\notag\\
&=\langle g,\tilde{g} |\mathcal{O}_1\cdots\mathcal{O}_nV|g,\tilde{g}\rangle.
\end{align}
Since the perturbation $V$ is quartic in the fermion fields, the expectation values of $V(T)$ between the thermal vacuum and odd-number excited states vanish. Therefore, $\langle0(\beta)|\alpha_{\bk}(T) V(T) |0(\beta)\rangle=\langle0(\beta)|\beta_{-\bk}(T) V(T) |0(\beta)\rangle=0$, and accordingly $V_{10}=0$.

The matrix element associated with the two-particle excited states, $V_{20}$, can be evaluated with the help of Eq.~(\ref{CTV2}) and the discussion below it.
Thus, the nonvanishing elements involve one $\alpha$- and one $\beta$- quanta and are given by
\begin{align}\label{V20}
V_{20,\mathbf{l}_1\bl_2}
=-2gu^*_{\bl_1}v^*_{\bl_1}\delta_{\bl_1,\bl_2}\sum_{\bq\neq \bl_1}v_{\bq}v^*_{\bq}.
\end{align}
To evaluate the matrix element associated with the four-particle excited states, $V_{40}$, we need to consider the following matrix elements
\begin{align}
&\langle0(\beta)|\beta_{-\bl_4}(T)\beta_{-\bl_3}(T)\beta_{-\bl_2}(T)\beta_{-\bl_1}(T) V(T) |0(\beta)\rangle,~\langle0(\beta)|\beta_{-\bl_4}(T)\beta_{-\bl_3}(T)\beta_{-\bl_2}(T)\alpha_{\bl_1}(T) V(T) |0(\beta)\rangle,\notag\\
&\langle0(\beta)|\beta_{-\bl_4}(T)\beta_{-\bl_3}(T)\alpha_{\bl_2}(T)\alpha_{\bl_1}(T) V(T) |0(\beta)\rangle,~\langle0(\beta)|\beta_{-\bl_4}(T)\alpha_{\bl_3}(T)\alpha_{\bl_2}(T)\alpha_{\bl_1}(T) V(T) |0(\beta)\rangle,\notag\\
&\langle0(\beta)|\alpha_{\bl_4}(T)\alpha_{\bl_3}(T)\alpha_{\bl_2}(T)\alpha_{\bl_1}(T) V(T) |0(\beta)\rangle.
\end{align}
It can be shown that only the term $\langle0(\beta)|\beta_{-\bl_4}(T)\beta_{-\bl_3}(T)\alpha_{\bl_2}(T)\alpha_{\bl_1}(T) V(T) |0(\beta)\rangle$ is nonzero (given by Eq.~(\ref{CTV4}) in the Appendix). Therefore, the nonvanishing matrix element is
\begin{align}\label{V40}
V_{40,\bl_1\bl_2\bl_3\bl_4}=&-g(u^*_{\bl_2}u^*_{\bl_4}v^*_{\bl_1}v^*_{\bl_3}+u^*_{\bl_1}u^*_{\bl_3}v^*_{\bl_2}v^*_{\bl_4})(1-\delta_{\bl_1,\bl_3})(1-\delta_{\bl_2,\bl_4})\delta_{\bl_1+\bl_2,\bl_3+\bl_4}\nonumber\\
&+g(u^*_{\bl_1}u^*_{\bl_4}v^*_{\bl_2}v^*_{\bl_3}+u^*_{\bl_2}u^*_{\bl_3}v^*_{\bl_1}v^*_{\bl_4})(1-\delta_{\bl_1,\bl_4})(1-\delta_{\bl_2,\bl_3})\delta_{\bl_1+\bl_2,\bl_3+\bl_4}.
\end{align}
Finally, since the perturbation $V$ contains at most four quasi-particle operators, all matrix elements associated with higher order ($k>4$) excited states vanish (see Eq.~\eqref{Vk0}).

According to Eq.~(\ref{CTVC}), the perturbed BCS thermal vacuum up to the first order is given by
\begin{align}\label{FTV}
|0(\beta)\rangle_\textrm{c}
&= |0(\beta)\rangle+\sum_{\bl_1,\bl_2}\alpha^\dag_{\bl_1}(T)\beta^\dag_{-\bl_2}(T)|0(\beta)\rangle\frac{V_{20,\bl_1\bl_2}}{E^{(0)}_0-E^{(0)}_{2,\bl_1\bl_2}} + \sum_{\bl_1,\bl_2,\bl_3,\bl_4}\alpha^\dag_{\bl_1}(T)\alpha^\dag_{\bl_2}(T)\beta^\dag_{-\bl_3}(T)\beta^\dag_{-\bl_4}(T)|0(\beta)\rangle\frac{V_{40,\bl_1\bl_2\bl_3\bl_4}}{E^{(0)}_0-E^{(0)}_{4,\bl_1\bl_2\bl_4\bl_4}}.
\end{align}
Here the matrix elements $V_{20}$ and $V_{40}$ are given by Eqs.~(\ref{V20}) and (\ref{V40}), respectively, $E^{(0)}_0=\sum_{\mathbf{k}}(\xi_{\mathbf{k}}-E_{\mathbf{k}})+\frac{|\Delta|^2}{g}$ is the unperturbed BCS ground-state energy, $E^{(0)}_{2,\bl_1\bl_2}=E^{(0)}_0+E_{\bl_1}+E_{\bl_2}$ with $E_{\bl_1}=\sqrt{\xi^2_{\bl_1}+|\Delta|^2}$ is the second-order excited state energy, and $E^{(0)}_{4,\bl_1\bl_2\bl_3\bl_4}=E^{(0)}_0+E_{\bl_1}+E_{\bl_2}+E_{\bl_3}+E_{\bl_4}$ is the fourth-order excited state energy.

After obtaining the expansion of of the full BCS thermal vacuum, one can derive the corrections to the chemical potential and gap function from the expectation values of the density and pairing operators. Explicitly,
\begin{align}
N=\sum_{\mathbf{k},\sigma=\uparrow,\downarrow} {}_\textrm{c}\langle0(\beta)|\psi_{\mathbf{k}\sigma}^\dagger\psi_{\mathbf{k}\sigma}|0(\beta)\rangle_\textrm{c},\quad
\Delta_\textrm{c}=-g\sum_\mathbf{k}{}_\textrm{c}\langle0(\beta)| \psi_{\mathbf{k}\uparrow}\psi_{\mathbf{-k}\downarrow}|0(\beta)\rangle_\textrm{c}.
\end{align}
It can be shown that the terms associated with $V_{40}$ do not contribute to $N$ or $\Delta_\textrm{c}$.
Following the convention of the BCS theory, we assume that the order parameter is a real number. We summarize the derivations in the Appendix and present here the final expressions. For the number equation, it becomes
\begin{align}\label{EoS2}
N=\sum_\bk\Big[1-\frac{\xi_\bk}{E_\bk}+2\frac{\xi_\bk}{E_\bk}f(E_\bk)\Big]+\sum_\bk\frac{1-f(E_\bk)}{2E^2_\bk}\frac{\Delta^2}{E^{(0)}_0-E^{(0)}_{2,\bk\bk}}\sum_{\bq\neq\bk}\Big(1-\frac{\xi_\bq}{E_\bq}\Big)+O(V^2),\notag\\
\end{align}
By solving the full chemical potential $\mu_c$ from the equation, one can obtain the correction to $\mu$.
The expansion of the order parameter can be obtained in a similar fashion:
\begin{align}\label{CD}
\Delta_\textrm{c}=\Delta-g^2\sum_\bk\frac{1-f(E_\bk)}{8E^3_\bk}\frac{\Delta^2\xi_\bk}{E_\bk}\sum_{\bq\neq\bk}\Big(1-\frac{\xi_\bq}{E_\bq}\Big)+O(V^2).
\end{align}
Here we emphasize that $\Delta_\textrm{c}$ ($\Delta$) denotes the full (unperturbed) order parameter.

\begin{figure}[t]
\centering
\includegraphics[width=4.4in, clip]{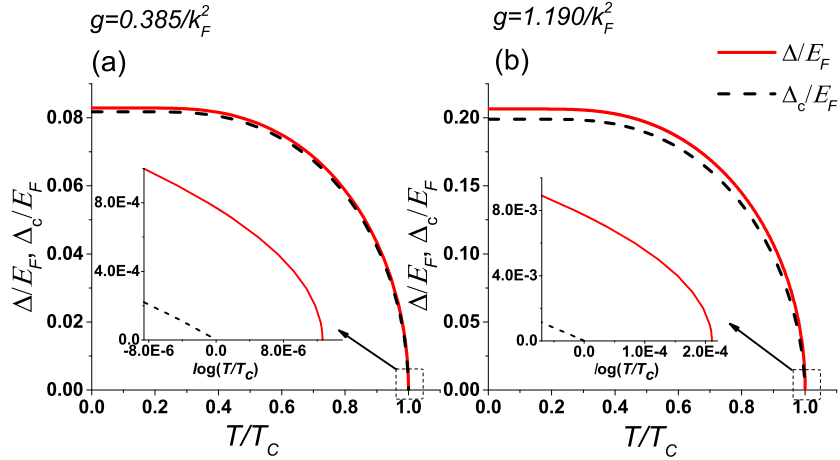}
 \caption{ The unperturbed (black dash lines) and perturbed (red solid lines) order parameters as functions of temperature for (a) $g=0.385/k_F^2$ and (b) $g=1.19/k_F^2$. The perturbation is truncated at the second order. The insets show the details of the curves near $T_c$, and there the temperature is shown in logarithmic scale. Here $E_F=\hbar^2k^2_F/(2m)$ is the Fermi energy of a noninteracting Fermi gas with the same density.}
 \label{fig.1}
\end{figure}

The first-order correction to the order parameter can be found numerically for different coupling strengths. The fermion density is fixed at $n=\frac{N}{\mathcal{V}}=\frac{k^3_F}{3\pi^2}$ with $k_F$ being the Fermi momentum of a noninteracting
Fermi gas with the same particle density. We first solve the unperturbed equations of states, Eqs. (\ref{Neqn}) and (\ref{Geqn}), at different temperatures to obtain $\Delta$ and $\mu$. Then, we substitute the unperturbed solution to Eq.~(\ref{CD}) and get the first-order correction to $\Delta_\textrm{c}(T)$. The critical temperature $T_c$ can be found by checking where the full order parameter $\Delta_\textrm{c}(T)$ vanishes. Eq.~(\ref{CD}) indicates the full order parameter is lowered by the first-order correction. We found that the critical temperature $T_c$ is also lowered when compared to the unperturbed value. However, our numerical results show that the correction to the critical temperature is small if the particle-particle interaction is weak. The strong correction to $T_c$ due to the particle-hole channel (induced interaction) \cite{GMB61,Heiselberg00} is not included in the BCS theory. Since the perturbation (\ref{pV}) considered here carries non-zero momentum, the calculation here shows the correction from finite-momentum effects to the Cooper pairs.

Figure~\ref{fig.1} shows the unperturbed and perturbed (up to the first-order correction) order parameters as functions of temperature. In Fig.~\ref{fig.1} (a), a relatively small coupling constant $g=0.385/k^2_F$ is chosen, which corresponds to the conventional BCS case where the interaction energy is much smaller compared to the Fermi energy. We found $\delta \Delta/\Delta\simeq 10^{-2}$ at any temperature below $T_c$ (which is determined by $\Delta_\textrm{c}(T_c)=0$), where $\delta\Delta=\Delta_\textrm{c}-\Delta$ is the first order correction of the order parameter. Hence, $\delta \Delta$ is indeed small in the BCS limit. We also found the ratio $\frac{\Delta_\textrm{c}(0)}{k_BT_c}\simeq1.72$, which is close to the mean-field BCS result of $1.76$~\cite{Tinkham_SCbook,Fetter_book}.
In Fig.~\ref{fig.1} (b), a relatively large coupling constant $g=1.19/k^2_F$ is chosen. The first-order correction $\delta \Delta/\Delta$ is more visible, but the ratio is still less than $8\%$ at any temperature below $T_c$. We also found $\frac{\Delta_\textrm{c}(0)}{k_BT_c}\simeq1.61$, which is more distinct from the unperturbed BCS value. According to the value $\Delta(T=0)/E_F$ indicated by Fig.~\ref{fig.1}, the system with $g=0.385/k^2_F$ is in the BCS regime while the one with $g=1.19/k^2_F$ is beyond the BCS limit because of its relatively large gap. 
The perturbation calculations allow us to improve the results order by order, but the complexity of the calculations increases rapidly. The insets of Fig.~\ref{fig.1} show the detailed behavior close to $T_c$ and indeed the critical temperature is lowered by the perturbation.

We remark that the BCS theory is usually viewed as a variational theory~\cite{BCS57,Tinkham_SCbook,Schrieffer_book}. The introduction of the perturbation calculation using the BCS thermal vacuum reproduces the thermodynamics of the BCS theory at the lowest order, and it introduces a quantum-mechanical style perturbation theory. The corrections to the BCS theory thus can be obtained by the perturbation theory.


\section{Applications of BCS thermal vacuum}\label{secV}
\subsection{Pairing Correlation}
In principle, the thermal vacuum provides a method for directly evaluating the statistical (thermal) average of any operator constructed from the fermion field $\psi$. Here we demonstrate another example by analyzing the correlation between the Cooper pairs, coming from higher moments of the order parameter. We introduce the pairing correlation
\begin{align}\label{pfdef2}\Delta^2_\textrm{p}=\langle V_p\rangle^2-\langle V_p^2\rangle,\end{align}
where
\begin{equation}
V_p=g\sum_{\bp} \psi_{\bp\uparrow} \psi_{-\bp\downarrow}
\end{equation}
is the pairing operator and $\langle V_p\rangle=\Delta$ (or $\Delta_\textrm{c}$ depending on whether the unperturbed or perturbed BCS thermal vacuum is used) if we choose the order parameter to be real. If the system obeys number conservation or the Wick decomposition~\cite{Walecka}, then $\Delta^2_\textrm{p}=0$ and the system exhibits no pairing correlation. A straightforward calculation shows that $\Delta^2_\textrm{p}\ge 0$ at the level of the unperturbed BCS thermal vacuum:
\begin{align}\label{pfdef3}
\Delta^2-\langle V_p^2\rangle_0=\Delta^2-\langle 0(\beta)|V_p^2|0(\beta)\rangle=g^2\sum_\bp \Big\{\frac{\Delta^* }{2E_\bp}\big[1-2f(E_{\bp})\big]\Big\}^2.
\end{align}
We remark that one may consider the pairing fluctuation by defining
$\tilde{\Delta}^2_\textrm{p}=\langle V_p^\dagger V_p\rangle-|\langle V_p\rangle|^2$.
However, this expression mixes both the pair-pair (e.g. $\psi^\dagger \psi^\dagger$ - $\psi\psi$) correlation as well as the density-density (e.g. $\psi^\dagger \psi$ -$\psi^\dagger \psi$) correlation and does not clearly reveal the effects from pairing. Therefore, we use the expression~\eqref{pfdef2} to investigate the pair-pair correlation.

Next, we use the perturbation theory to evaluate higher-order corrections to the pairing correlation. We take the perturbative BCS thermal vacuum shown in Eq.~(\ref{FTV}) and estimate the correction to the pairing correlation:
\begin{align} \label{310}
\Delta^2_\textrm{p}&=\Delta_\textrm{c}^2-{}_\textrm{c}\langle0(\beta)| V_p^2|0(\beta)\rangle_\textrm{c}\notag \\
&=-g^2\sum_\bp(u_\bp v_\bp \sin^2\theta_\bp-u_\bp v_\bp\cos^2\theta_\bp)^2+g^2\big(\sum_\bp u_\bp v_\bp\cos2\theta_\bp\big)^2 \notag\\&+\sum_{\bl_1} [2g^2u^2_{\bl_1}\cos^2\theta_{\bl_1}\Delta -2g^3u^3_{\bl_1}v_{\bl_1}\cos^2\theta_{\bl_1}\cos 2\theta_{\bl_1}]\dfrac{u_{\bl_1}v_{\bl_1}\sum_{\bq\neq \bl_1}v_{\bq}v_{\bq}}{E_{\bl_1}} \notag \\
&-\sum_{\bl_1} [2g^2v^{2}_{\bl_1}\cos^2\theta_{\bl_1}\Delta-2g^3u_{\bl_1}v^{3}_{\bl_1}\cos^2\theta_{\bl_1}\cos 2\theta_{\bl_1}]\dfrac{u_{\bl_1}v_{\bl_1}\sum_{\bq\neq \bl_1}v_{\bq}v_{\bq}}{E_{\bl_1}}\notag\\& +\sum_{\bl_1\neq\bl_2} 4g^3u^2_{\bl_1}u^2_{\bl_2}\cos^2\theta_{\bl_1}\cos^2\theta_{\bl_2}\dfrac{u_{\bl_2}v_{\bl_2}u_{\bl_1}v_{\bl_1}}{(E_{\bl_1}+E_{\bl_2})}  +\sum_{\bl_1\neq\bl_2} 4g^3v^{2}_{\bl_1}v^{2}_{\bl_2}\cos^2\theta_{\bl_1}\cos^2\theta_{\bl_2}\dfrac{u_{\bl_2}v_{\bl_2}u_{\bl_1}v_{\bl_1}}{(E_{\bl_1}+E_{\bl_2})}-\Delta_\textrm{c}^2. 
\end{align}
The expression of ${}_\textrm{c}\langle0(\beta)| V_p^2|0(\beta)\rangle_\textrm{c}$ to the first order is shown in Eq.~\eqref{313} in the Appendix.

\begin{figure}[th]
	\centering
	\includegraphics[width=4.4in, clip]{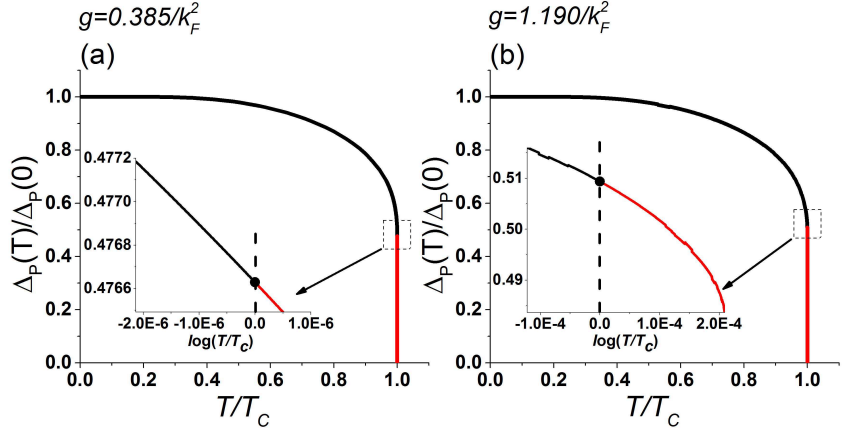}
	\caption{The pairing correlation $\Delta_\textrm{p}(T)$, up to the first order according to Eq.~\eqref{310}, as a function of temperature (normalized by $\Delta_\textrm{p}(T=0)$) for (a) $g=0.385/k^2_F$ and (b) $g=1.190/k^2_F$, respectively. The insets show the behavior close to $T_c$ (indicated by the black dots). The curves above and below $T_c$ are colored by black and red to emphasize the pseudogap (red) effect. }
	\label{fig.3}
\end{figure}

Figure \ref{fig.3} shows the pairing fluctuation $\Delta_\textrm{p}$ up to the first order according to Eq.~(\ref{310}) as functions of the temperature.
The pairing fluctuation decreases with temperature but increases with the interaction.
By examining $\Delta_\textrm{p}$ near $T_c$, we found that the pairing correlation survives in a small region above the corrected $T_c$. In other words, when the corrected order parameter $\Delta_c$ vanishes, the pairing correlation persists. This may be considered as evidence of the pseudogap effect~\cite{Ourreview,OurAnnPhys}, where pairing effects still influence the system above $T_c$. We mention that in the Ginzburg-Landau theory, the Ginzburg criterion checks the fluctuation of the specific heat~\cite{Mazenko_book} to identify the critical regime. Here we directly evaluate the pairing correlation by applying the BCS thermal vacuum and its perturbation theory by estimating the correction from the original fermion-fermion interaction. Our results provide a direct calculation the pairing correlation and offer support for the pseudogap phenomenon.

\subsection{Generalized Squeezed Coherent State}
The unperturbed BCS thermal vacuum itself has some interesting properties. Since the BCS thermal vacuum is obtained from a Bogoliubov transformation, it is a generalized coherent state.
To verify the conjecture, we follow Ref.~\cite{OurPLA17} and introduce the temperature-independent spin operators
\begin{align}
&S_{\alpha \mathbf{k}}^+=\alpha_{\mathbf{k}}^\dagger\tilde{\alpha}_{\mathbf{k}}^\dagger,\quad S_{\alpha \mathbf{k}}^-=(S_{\alpha \mathbf{k}}^+)^\dagger=\tilde{\alpha}_{\mathbf{k}}\alpha_{\mathbf{k}},\quad S_{\alpha \mathbf{k}}^z=\frac{1}{2}[S_{\alpha \mathbf{k}}^+,S_{\alpha \mathbf{k}}^-]=\frac{1}{2}(\alpha_{\mathbf{k}}^\dagger\alpha_{\mathbf{k}}+\tilde{\alpha}_{\mathbf{k}}^\dagger\tilde{\alpha}_{\mathbf{k}}-1),\notag \\
&S_{\beta \mathbf{k}}^+=\beta_{\mathbf{-k}}^\dagger\tilde{\beta}_{-\mathbf{k}}^\dagger,\quad
S_{\beta \mathbf{k}}^-=(S_{\beta \mathbf{k}}^+)^\dagger=\tilde{\beta}_{-\mathbf{k}}\beta_{\mathbf{-k}},\quad S_{\beta \mathbf{k}}^z=\frac{1}{2}[S_{\beta \mathbf{k}}^+,S_{\beta \mathbf{k}}^-]=\frac{1}{2}(\beta_{\mathbf{-k}}^\dagger\beta_{\mathbf{-k}}+\tilde{\beta}_{-\mathbf{k}}^\dagger\tilde{\beta}_{-\mathbf{k}}-1),
\end{align}
which satisfy the SU(2) algebra since $[S_{\alpha \mathbf{k}}^z,S_{\alpha \mathbf{k}}^\pm]=\pm S_{\alpha \mathbf{k}}^\pm$ and $[S_{\beta \mathbf{k}}^z,S_{\beta \mathbf{k}}^\pm]=\pm S_{\beta \mathbf{k}}^\pm$.
The BCS thermal vacuum can be rewritten as
\begin{align}
\vert 0(\beta)\rangle=\mathrm{e}^{\sum_{\mathbf{k}} (\theta_\mathbf{k}\mathrm{e}^{-i\chi}S_{\alpha \mathbf{k}}^+-\theta_\mathbf{k}\mathrm{e}^{i\chi}S_{\alpha \mathbf{k}}^-)}\mathrm{e}^{\sum_{\mathbf{k}} (\theta_\mathbf{k}\mathrm{e}^{-i\chi}S_{\beta \mathbf{k}}^+-\theta_\mathbf{k}\mathrm{e}^{i\chi}S_{\beta \mathbf{k}}^-)}\vert g,\tilde{g}\rangle.
\end{align}
Since $S_{\alpha \mathbf{k}}^-\vert g,\tilde{g}\rangle=S_{\beta \mathbf{k}}^-\vert g,\tilde{g}\rangle=0$, the above expression shows that the BCS thermal vacuum is indeed a generalized SU(2) coherent state~\cite{OurPLA17}. However, it has another important property: The BCS thermal vacuum is a nilpotent coherent state because $(S_{\beta \mathbf{k}}^+)^2=0=(S_{\beta \mathbf{k}}^-)^2$. The identity should be understood as an identity in the Fock space of the quasi-particles.
However, it is important to notice that the BCS thermal vacuum is not a coherent state with respect to the $\psi_{\sigma}$-quanta.

Moreover, the BCS thermal vacuum is a squeezed state associated with $S_{\alpha \mathbf{k}}^{x,y}$, $S_{\beta \mathbf{k}}^{x,y}$. Here
\begin{eqnarray}
S_{\alpha \mathbf{k}}^x=\frac{1}{2}(S_{\alpha \mathbf{k}}^++S_{\alpha \mathbf{k}}^-),\quad
S_{\alpha \mathbf{k}}^y=\frac{1}{2i}(S_{\alpha \mathbf{k}}^+-S_{\alpha \mathbf{k}}^-),\quad S_{\beta \mathbf{k}}^x=\frac{1}{2}(S_{\beta \mathbf{k}}^++S_{\beta \mathbf{k}}^-),\quad
S_{\beta \mathbf{k}}^y=\frac{1}{2i}(S_{\beta \mathbf{k}}^+-S_{\beta \mathbf{k}}^-).
\end{eqnarray}
That means the BCS thermal vacuum saturates the the Robertson-Schrodinger inequality~\cite{Merzbacher_book}
\begin{equation}\label{RSi}
\sigma_A^2\sigma_B^2\geq\vert \frac{1}{2}\langle\lbrace A,B\rbrace\rangle-\langle A\rangle\langle B\rangle\vert^2+\vert \frac{1}{2i}\langle[A,B]\rangle\vert^2,
\end{equation}
where $A=S_{\alpha,\beta \mathbf{k}}^x$, $B=S_{\alpha,\beta \mathbf{k}}^y$, $\{A,B\}$ and $[A,B]$ denote the anti-commutator and commutator of $A$ and $B$.
The proof that the BCS thermal vacuum leads to an equal sign in the above inequality is summarized in the Appendix. Therefore, the BCS thermal vacuum is a squeezed coherent state.

\subsection{Geometric Phase of BCS Thermal Vacuum}
In the derivation of the unperturbed BCS thermal vacuum, we introduced a phase $\chi$ (see Eq.(\ref{ct212_3})).
Since $|0(\beta)\rangle$ may be considered as a pure quantum state, the system can acquire a geometric phase similar to the Berry phase~\cite{Berry84} when $\chi$ evolves adiabatically along a closed loop $C$ in the parameter space. We call it the thermal phase $\gamma(C)$, which can be evaluated as follows.
\begin{eqnarray}
\gamma(C)=i\oint_C dt\langle 0(\beta),\chi(t)\vert \frac{d}{dt}\vert 0(\beta),\chi(t)\rangle=i\int_0^{2\pi}d\chi \langle 0(\beta),\chi\vert \frac{\partial}{\partial \chi}\vert 0(\beta),\chi\rangle.
\end{eqnarray}
By using the Baker-Campbell-Hausdorff disentangling formula\cite{Campbell1897,Baker02}, the thermal vacuum can be written as
\begin{equation}
\vert 0(\beta)\rangle=\mathrm{e}^{\sum_{\mathbf{k}} 2\ln \cos \theta_{\mathbf{k}}}\mathrm{e}^{\sum_{\mathbf{k}} \tan\theta_{\mathbf{k}}\mathrm{e}^{-i\chi}(\alpha_{\mathbf{k}}^\dagger\tilde{\alpha}_{\mathbf{k}}^\dagger+\beta_{-\mathbf{k}}^\dagger\tilde{\beta}_{\mathbf{-k}}^\dagger)}\vert g,\tilde{g}\rangle.
\end{equation}
After some algebra, we obtain
$i\langle 0(\beta)\vert \frac{\partial}{\partial\chi}\vert 0(\beta)\rangle
=2\sum_{\mathbf{k}} f(E_{\mathbf{k}})$.
Therefore, the thermal phase is
\begin{eqnarray}\label{BerryBCS0}
\gamma(C)=4\pi\sum_{\mathbf{k}} f(E_{\mathbf{k}})=2\pi\big[N_\alpha(\beta)+N_\beta(\beta)\big],
\end{eqnarray}
which is proportional to the total quasi-particle number $N_\alpha(\beta)+N_\beta(\beta)$ at temperature $ T=\frac{1}{k_B\beta}$. The quasi-particle number is nonzero only when $T>0$.

The origin of the thermal phase can be understood as follows. The thermal vacuum can be thought of as a purification of the mixed state at finite temperatures. This can be clarified by noting that Eq.~(\ref{GDM}) leads to
\begin{eqnarray}\label{U1}
|0(\beta)\rangle&=&\sum_n\frac{1}{\sqrt{Z}}\textrm{e}^{-\frac{\beta E_n}{2}-in\chi}|n,\tilde{n}\rangle
=\sqrt{\rho_\beta}\sum_n\mathcal{U}|n,\tilde{n}\rangle,
\end{eqnarray}
where $\rho_\beta=\frac{1}{Z}\textrm{e}^{-\beta H}$ is the density matrix of the non-tilde system at temperature $T$, and the phase $n\chi$ is because each excitation of a quasi-particle contributes $\me^{-i\chi}$ indicated by Eq.~(\ref{ct212_3}). The unitary operator $\mathcal{U}$ has the following matrix representation in the basis formed by $\{|n\rangle\}$:
\begin{eqnarray}\label{U2}
\left(\begin{array}{cccc}
1 & 0 & 0 & \cdots \\
0 & e^{-i\chi} & 0 &  \cdots \\
0 & 0 & e^{-2i\chi} & \cdots \\
\vdots & \vdots & \vdots & \ddots \end{array}\right).
\end{eqnarray}
There is another way of purifying the density matrix~\cite{UPPRL14,GPhase_book} by defining the amplitude of the density matrix as
\begin{eqnarray}\label{w1}
w_\beta&=&\sqrt{\rho_\beta}\mathcal{U}
=\sum_n\frac{1}{\sqrt{Z}}\textrm{e}^{-\frac{\beta H}{2}}|n\rangle\langle n|\textrm{e}^{-in\chi}.
\end{eqnarray}
By comparing Eqs.~(\ref{U1}) and (\ref{w1}), one can find a one-to-one mapping between the thermal vacuum $|0(\beta)\rangle$ and the amplitude $w_{\beta}$.

However, the purification is not unique. For instance, Eq.~(\ref{f0f1}) allows a relative phase $\chi$ between $f_{0\mathbf{k}}$ and $f_{1\mathbf{k}}$. A U(1) transformation corresponding to a change of the parameter $\chi$ leads to another thermal vacuum. Hence, the BCS thermal vacuum can be parametrized in the space $0\le \chi < 2\pi$, i.e. one may recognize the collection of BCS thermal vacua as a U(1) manifold  parametrized by $\chi$.
The thermal phase (\ref{BerryBCS0}) from the BCS thermal vacuum may be understood as follows. If $\chi$ is transported along the U(1) manifold along a loop, every excited quasi-particle acquires a phase $2\pi$. Therefore, the thermal phase indicates the number of thermal excitations in the system.

When $T\rightarrow 0$, the statistical average becomes the expectation value with respect to the ground state. In the present case, the BCS thermal vacuum reduces to the (two-mode) BCS ground state. As a consequence, the U(1) manifold of the unitary transformation of the BCS thermal vacua is no longer defined at $T= 0$ because there is no thermal excitation and $f_{1\bk}=0$ in Eq.~\eqref{eq:f0f1}. Importantly, the BCS ground state is already a pure state at $T=0$, so there is no need to introduce $\chi$ for parametrizing the manifold of the unitary transformation in the purification. Hence, the thermal phase should only be defined at finite temperatures when the system is thermal.

\section{Conclusion}\label{conclusion}
By introducing the two-mode BCS vacuum and the corresponding unitary transformation, we have shown how to construct the BCS thermal vacuum. A perturbation theory is then developed based on the BCS thermal vacuum. In principle, one can evaluate the corrections from the original fermion interactions ignored in the BCS approximation. Importantly, the perturbation calculations are at the quantum-mechanical level even though the BCS theory and the BCS thermal vacuum are based on quantum field theory.

The BCS thermal vacuum is expected to offer more insights into interacting quantum many-body systems. We have shown that the pairing correlation from the perturbation theory persists when the corrected order parameter vanishes, offering evidence of the pseudogap phenomenon. In addition to the saturation of the Robertson-Schrodinger inequality by the BCS thermal vacuum, the thermal phases associated with the BCS thermal vacuum elucidates the internal geometry of its construction. The BCS thermal vacuum and its perturbation theory thus offers an alternative way for investigating superconductivity and superfluidity.

\textit{Acknowledgment}: We thank Fred Cooper for stimulating discussions. H. G. thanks the support from the National Natural Science Foundation of China (Grant No. 11674051).

\appendix
\section{Details of perturbation theory based on BCS thermal vacuum}
Here are some details of the calculations involving the BCS thermal vacuum.
For the calculations of the matrix elements $V_{20,\bl_1\bl_2}$ in the perturbation theory, we evaluate
\begin{eqnarray}\label{CTV2}
& &\langle 0(\beta)|\beta_{-\bl_2}(T)\alpha_{\bl_1}(T) V(T) |0(\beta)\rangle
=-g\sum_{\bk\neq 0}\sum_{\bp\bq} \langle g,\tilde{g}|\me^{-Q}\beta_{-\bl_2}(T)\alpha_{\bl_1}(T) \me^{Q}(u^*_{\bk-\bp}\alpha^\dag_{\bk-\bp}+v_{\bk-\bp}\beta_{\bp-\bk})\nonumber \\
&\times&(u^*_{-\bp}\beta^\dag_{\bp}-v_{-\bp}\alpha_{-\bp})(u_{\bq-\bk}\beta_{\bk-\bq}-v^*_{\bq-\bk}\alpha^\dag_{\bq-\bk})(u_\bq\alpha_{\bq}+v^*_\bq\beta_{-\bq}^\dag) \me^{-Q}\me^Q|g,\tilde{g}\rangle \nonumber \\
&=&-g\sum_{\bk\neq 0}\sum_{\bp\bq} (u^*_{\bk-\bp}u^*_{-\bp}u_{\bq-\bk}v^*_\bq
\delta_{\bl_1,\bk-\bp}\delta_{-\bl_2,\bp}\delta_{\bk-\bq,-\bq} \nonumber \\
&+&u^*_{\bk-\bp}v_{-\bp}v^*_{\bq-\bk}v^*_\bq
\delta_{-\bp,\bq-\bk}\delta_{\bl_1,\bk-\bp}\delta_{-\bl_2,-\bq}
-v_{\bk-\bp}u^*_{-\bp}v^*_{\bq-\bk}v^*_\bq
\delta_{\bl_1,\bq-\bk}\delta_{\bp-\bk,\bp}\delta_{-\bl_2,-\bq} \nonumber \\
&+&v_{\bk-\bp}u^*_{-\bp}v^*_{\bq-\bk}v^*_\bq
\delta_{\bl_1,\bq-\bk}\delta_{-\bl_2,\bp}\delta_{\bp-\bk,-\bq}) \nonumber \\
&=&-2gu^*_{\bl_1}v^*_{\bl_1}\delta_{\bl_1,\bl_2}\sum_{\bq\neq \bl_1}v_{\bq}v^*_{\bq},
\end{eqnarray}
where we have used $0=\alpha_\bk|g,\tilde{g}\rangle=\langle g,\tilde{g}|\alpha^\dag_\bk=\beta_{-\bk}|g,\tilde{g}\rangle=\langle g,\tilde{g}|\beta^\dag_{-\bk}$, and $|u_\bk|^2,|v_\bk|^2=(1\pm\xi_\bk/E_\bk)/2$.
Similar calculations lead to $\langle 0(\beta)|\alpha_{\bl_2}(T)\alpha_{\bl_1}(T) V(T) |0(\beta)\rangle=0=\langle 0(\beta)|\beta_{-\bl_2}(T)\beta_{-\bl_1}(T) V(T) |0(\beta)\rangle$. Therefore, $V_{20,\bl_1\bl_2}=V_{20,\bl_1}\delta_{\bl_1,\bl_2}$ with $V_{20,\bl_1}=-2gu^*_{\bl_1}v^*_{\bl_1}\sum_{\bq\neq \bl_1}v_{\bq}v^*_{\bq}$.

Next, we evaluate $V_{40,\bl_1\bl_2\bl_3\bl_4}$:
\begin{align}\label{CTV4}
& \langle 0(\beta)|\beta_{-\bl_4}(T)\beta_{-\bl_3}(T)\alpha_{\bl_2}(T)\alpha_{\bl_1}(T) V(T) |0(\beta)\rangle=-\sum_{\bk\neq 0}\sum_{\bp\bq} \langle g,\tilde{g}|\beta_{-\bl_4}\beta_{-\bl_3}\alpha_{\bl_2}\alpha_{\bl_1}V| g,\tilde{g}\rangle\nonumber \\
=&-g(u^*_{\bl_2}u^*_{\bl_4}v^*_{\bl_1}v^*_{\bl_3}+u^*_{\bl_1}u^*_{\bl_3}v^*_{\bl_2}v^*_{\bl_4})(1-\delta_{\bl_1,\bl_3})(1-\delta_{\bl_2,\bl_4})\delta_{\bl_1+\bl_2,\bl_3+\bl_4}\notag\\&
+g(u^*_{\bl_1}u^*_{\bl_4}v^*_{\bl_2}v^*_{\bl_3}+u^*_{\bl_2}u^*_{\bl_3}v^*_{\bl_1}v^*_{\bl_4})(1-\delta_{\bl_1,\bl_4})(1-\delta_{\bl_2,\bl_3})\delta_{\bl_1+\bl_2,\bl_3+\bl_4}.
\end{align}
Hence,
\begin{align}
V_{40,\bl_1\bl_2\bl_3\bl_4}=&-g(u^*_{\bl_2}u^*_{\bl_4}v^*_{\bl_1}v^*_{\bl_3}+u^*_{\bl_1}u^*_{\bl_3}v^*_{\bl_2}v^*_{\bl_4})(1-\delta_{\bl_1,\bl_3})(1-\delta_{\bl_2,\bl_4})\delta_{\bl_1+\bl_2,\bl_3+\bl_4}\notag\\&+g(u^*_{\bl_1}u^*_{\bl_4}v^*_{\bl_2}v^*_{\bl_3}+u^*_{\bl_2}u^*_{\bl_3}v^*_{\bl_1}v^*_{\bl_4})(1-\delta_{\bl_1,\bl_4})(1-\delta_{\bl_2,\bl_3})\delta_{\bl_1+\bl_2,\bl_3+\bl_4}.
\end{align}

The total particle number is given by the expectation value of the $\psi$-quantum number operator with respect to the state $|0(\beta)\rangle_\textrm{c}$:
\begin{align} \label{31}
N&=\sum_{\bk,\sigma=\uparrow,\downarrow}{}_\textrm{c}\langle0(\beta)|\psi_{\mathbf{k}\sigma}^\dagger\psi_{\mathbf{k}\sigma}|0(\beta)\rangle_\textrm{c}
= \sum_\bk\langle 0(\beta)|\big(\psi_{\mathbf{k}\uparrow}^\dagger\psi_{\mathbf{k}\uparrow} + \psi_{-\mathbf{k}\downarrow}^\dagger\psi_{-\mathbf{k}\downarrow}\big)|0(\beta)\rangle \notag\\&
+\sum_{\bl_3}\sum_{\bl_1,\bl_2} \big(\langle0(\beta)|\psi_{\bl_3\uparrow}^\dagger\psi_{\bl_3\uparrow}\alpha^\dag_{\bl_1}(T)\beta^\dag_{-\bl_2}(T)|0(\beta)\rangle\frac{V_{20,\mathbf{l}_1\bl_2}}{E^{(0)}_0-E^{(0)}_{2,\bl_1\bl_2}} +\langle 0(\beta)| \psi_{-\bl_3\downarrow}^\dagger\psi_{-\bl_3\downarrow}\alpha^\dag_{\bl_1}(T)\beta^\dag_{-\bl_2}(T)|0(\beta)\rangle\frac{V_{20,\mathbf{l}_1\bl_2}}{E^{(0)}_0-E^{(0)}_{2,\bl_1\bl_2}}\big) \notag\\&+\sum_{\bl_3}\sum_{\bl_1,\bl_2} \big(\langle0(\beta)|\beta_{-\bl_2}(T)\alpha_{\bl_1}(T)\psi_{\bl_3\uparrow}^\dagger\psi_{\bl_3\uparrow}|0(\beta)\rangle
\frac{V^*_{20,\mathbf{l}_1\bl_2}}{E^{(0)}_0-E^{(0)}_{2,\bl_1\bl_2}} +\langle 0(\beta)|\beta_{-\bl_2}(T)\alpha_{\bl_1}(T) \psi_{-\bl_3\downarrow}^\dagger\psi_{-\bl_3\downarrow}|0(\beta)\rangle\frac{V^*_{20,\mathbf{l}_1\bl_2}}{E^{(0)}_0-E^{(0)}_{2,\bl_1\bl_2}}\big)\notag\\&+O(V^2) \notag \\
&=\sum_\bk\Big(1-\frac{\xi_\bk}{E_\bk}+2\frac{\xi_\bk}{E_\bk}f(E_\bk)\Big) +\sum_\bk\frac{1-f(E_\bk)}{E_\bk}\frac{\Delta V_{20,\bk}+\Delta^*V^*_{20,\bk}}{E^{(0)}_0-E^{(0)}_{2,\bk\bk}}+O(V^2).
\end{align}
The perturbation series of the gap function can be derived in a similar fashion.

To evaluate the pairing correlation, we calculate the unperturbed expectation
\begin{eqnarray}\label{20}
\langle0(\beta)|V^2_p|0(\beta)\rangle
&=&g^2\sum_{\bp\bq}\langle g,\tilde{g}|\me^{-Q}(u_{\bp}\alpha_{\bp}+v^*_{\bp}\beta_{-\bp}^\dag)(u_{\bp}\beta_{-\bp}-v^*_{\bp}\alpha^\dag_{\bp})
(u_\bq\alpha_{-\bq}+v^*_\bq\beta_{\bq}^\dag)(u_{\bq}\beta_{\bq}-v^*_{\bq}\alpha^\dag_{-\bq}) \me^Q|g,\tilde{g}\rangle\notag\\
&=&-g^2\sum_\bp(u_\bp v_\bp^* \sin^2\theta_\bp-u_\bp v^*_\bp\cos^2\theta_\bp)^2+g^2\big(\sum_\bp u^*_\bp v_\bp\cos2\theta_\bp\big)^2.
\end{eqnarray}
By using the BCS thermal vacuum, we obtain the following expression to the first order.
\begin{align}
\langle V^2_p\rangle&\equiv {}_\textrm{c}\langle 0(\beta)|V^2_p|0(\beta)\rangle_\textrm{c} \notag \\
&=\langle 0(\beta)|V^2_p|0(\beta)\rangle \notag\\
&+\sum_{\bl_1,\bl_2} \big(\langle0(\beta)|V^2_p\alpha^\dag_{\bl_1}(T)\beta^\dag_{-\bl_2}(T)|0(\beta)\rangle\frac{V_{20,\mathbf{l}_1\bl_2}}{E^{(0)}_0-E^{(0)}_{2,\bl_1\bl_2}}
+\sum_{\bl_1,\bl_2} \big(\langle0(\beta)|\beta_{-\bl_2}(T)\alpha_{\bl_1}(T)V^2_p|0(\beta)\rangle\frac{V^*_{20,\mathbf{l}_1\bl_2}}{E^{(0)}_0-E^{(0)}_{2,\bl_1\bl_2}} \notag \\
&+\sum_{\bl_1,\bl_2,\bl_3,\bl_4} \frac{V_{40,\mathbf{l}_1\bl_2\bl_3\bl_4}}{E^{(0)}_0-E^{(0)}_{4,\bl_1\bl_2\bl_3\bl_4}} \big(\langle0(\beta)|V^2_p\alpha^\dag_{\bl_1}(T)\alpha^\dag_{\bl_2}(T)\beta^\dag_{-\bl_3}(T)\beta^\dag_{-\bl_4}(T)|0(\beta)\rangle
\notag\\&+\sum_{\bl_1,\bl_2,\bl_3,\bl_4} \frac{V^*_{40,\mathbf{l}_1\bl_2\bl_3\bl_4}}{E^{(0)}_0-E^{(0)}_{4,\bl_1\bl_2\bl_3\bl_4}}\big(\langle0(\beta)|\beta_{-\bl_4}(T)\beta_{-\bl_3}(T)\alpha_{\bl_2}(T)\alpha_{\bl_1}(T)V^2_p|0(\beta)\rangle. \notag \\
&=\langle V^2_p\rangle_0 +\sum_{\bl_1,\bl_2} \Big(\frac{V_{p2,\bl_1\bl_2}V_{20,\mathbf{l}_1\bl_2}}{E^{(0)}_0-E^{(0)}_{2,\bl_1\bl_2}} +\frac{V^*_{p2,\bl_1\bl_2}V^*_{20,\mathbf{l}_1\bl_2}}{E^{(0)}_0-E^{(0)}_{2,\bl_1\bl_2}}\Big)+\sum_{\bl_1,\bl_2,\bl_3,\bl_4} \Big(\frac{V_{p4,\bl_1\bl_2\bl_3\bl_4}V_{40,\mathbf{l}_1\bl_2\bl_3\bl_4}}{E^{(0)}_0-E^{(0)}_{4,\bl_1\bl_2\bl_3\bl_4}}
+\frac{V^*_{p4,\bl_1\bl_2\bl_3\bl_4}V^*_{40,\mathbf{l}_1\bl_2\bl_3\bl_4}}{E^{(0)}_0-E^{(0)}_{4,\bl_1\bl_2\bl_3\bl_4}}\Big),  \label{312}
\end{align}
where 
\begin{align}
&V_{p2,\bl_1\bl_2}=\langle0(\beta)|V^2_p\alpha^\dag_{\bl_1}(T)\beta^\dag_{-\bl_2}(T)|0(\beta)\rangle,\quad V^*_{p2,\bl_1\bl_2}=\langle0(\beta)|\beta_{-\bl_2}(T)\alpha_{\bl_1}(T)(V^\dag_p)^2|0(\beta)\rangle,\notag\\
&V_{p4,\bl_1\bl_2\bl_3\bl_4}=\langle0(\beta)|V^2_p\alpha^\dag_{\bl_1}(T)\alpha^\dag_{\bl_2}(T)\beta^\dag_{-\bl_3}(T)\beta^\dag_{-\bl_4}(T)|0(\beta)\rangle,~V^*_{p4,\bl_1\bl_2\bl_3\bl_4}=\langle0(\beta)|\beta_{-\bl_4}(T)\beta_{-\bl_3}(T)\alpha_{\bl_2}(T)\alpha_{\bl_1}(T)(V^\dag_p)^2|0(\beta)\rangle.
 \end{align}
 We remark that $V^\dag_p\neq V_p$.
The coefficients $V_{p2}$ and $V_{p4}$ are evaluated as follows.
\begin{widetext}
\begin{eqnarray}
V_{p2,\bl_1\bl_2}
&=&[2gu^2_{\bl_1}\cos^2\theta_{\bl_1}\Delta^*-2g^2u^3_{\bl_1}v^{*}_{\bl_1}\cos^2\theta_{\bl_1}\cos 2\theta_{\bl_1}]\delta_{\bl_1,\bl_2}
,  \label{24} \\
V^*_{p2,\bl_1\bl_2}
&=&[-2gv^{*2}_{\bl_1}\cos^2\theta_{\bl_1}\Delta^*+2g^2u_{\bl_1}v^{*3}_{\bl_1}\cos^2\theta_{\bl_1}\cos 2\theta_{\bl_1}]\delta_{\bl_1,\bl_2}
, \label{25} \\
V_{p4,\bl_1\bl_2\bl_3\bl_4}
&=&-2g^2u^2_{\bl_1}u^2_{\bl_2}\cos^2\theta_{\bl_1}\cos^2\theta_{\bl_2}(\delta_{\bl_1,\bl_3}\delta_{\bl_2,\bl_4}-\delta_{\bl_1,\bl_4}\delta_{\bl_2,\bl_3}), \label{26} \\
V^*_{p4,\bl_1\bl_2\bl_3\bl_4}
&=&-2g^2v^{*2}_{\bl_1}v^{*2}_{\bl_2}\cos^2\theta_{\bl_1}\cos^2\theta_{\bl_2}(\delta_{\bl_1,\bl_3}\delta_{\bl_2,\bl_4}-\delta_{\bl_1,\bl_4}\delta_{\bl_2,\bl_3}).\label{27}
\end{eqnarray}
Therefore, Eq.~(\ref{312}) becomes
\begin{eqnarray}\label{313}
\langle V^2_p\rangle
&=&\langle V^2_p\rangle_0
+\sum_{\bl_1} [2g^2u^2_{\bl_1}\cos^2\theta_{\bl_1}\Delta^*-2g^3u^3_{\bl_1}v^{*}_{\bl_1}\cos^2\theta_{\bl_1}\cos 2\theta_{\bl_1}]\dfrac{u^*_{\bl_1}v^*_{\bl_1}\sum_{\bq\neq \bl_1}v_{\bq}v^*_{\bq}}{E_{\bl_1}} \notag \\
&-&\sum_{\bl_1} [2g^2v^{*2}_{\bl_1}\cos^2\theta_{\bl_1}\Delta^*-2g^3u_{\bl_1}v^{*3}_{\bl_1}\cos^2\theta_{\bl_1}\cos 2\theta_{\bl_1}]\dfrac{u_{\bl_1}v_{\bl_1}\sum_{\bq\neq \bl_1}v^*_{\bq}v_{\bq}}{E_{\bl_1}} \notag \\
&+&\sum_{\bl_1\neq\bl_2} 4g^3u^2_{\bl_1}u^2_{\bl_2}\cos^2\theta_{\bl_1}\cos^2\theta_{\bl_2}\dfrac{u^*_{\bl_2}v^*_{\bl_2}u^*_{\bl_1}v^*_{\bl_1}}{(E_{\bl_1}+E_{\bl_2})} +\sum_{\bl_1\neq\bl_2} 4g^3v^{*2}_{\bl_1}v^{*2}_{\bl_2}\cos^2\theta_{\bl_1}\cos^2\theta_{\bl_2}\dfrac{u_{\bl_2}v_{\bl_2}u_{\bl_1}v_{\bl_1}}{(E_{\bl_1}+E_{\bl_2})},
\end{eqnarray}
\end{widetext}
where $\langle V^2_p\rangle_0$ is given by Eq.(\ref{20}).

\section{Proof of BCS thermal vacuum being a squeeze state}
By applying the relation $[S_{\alpha,\beta \mathbf{k}}^x,S_{\alpha,\beta \mathbf{k}}^y]=iS_{\alpha,\beta \mathbf{k}}^z$ for the temperature-independent spin operators, the proof of the saturation of the Robertson-Schrodinger inequality is equivalent to the proof of
\begin{align}
\label{ct242_1}
(\langle S^{x2}_{\beta \mathbf{k}}\rangle-\langle S_{\beta \mathbf{k}}^x\rangle^2)(\langle S^{y2}_{\beta \mathbf{k}}\rangle-\langle S_{\beta \mathbf{k}}^y\rangle^2)=(\langle S_{\beta \mathbf{k}}^x\rangle \langle S_{\beta \mathbf{k}}^y\rangle)^2+\frac{1}{4}\langle S_{\beta \mathbf{k}}^z\rangle^2
\end{align}
with respect to the BCS thermal vacuum $|0(\beta)\rangle$.
With the help of Eqs.~(\ref{eq:BogoT}), we have
\begin{align}
\langle S_{\beta \mathbf{k}}^x\rangle&=\frac{1}{2}\langle \beta_{\mathbf{-k}}^\dagger\tilde{\beta}^\dagger_{-\mathbf{k}}\rangle+\frac{1}{2}\langle \tilde{\beta}_{-\mathbf{k}}\beta_{-\mathbf{k}}\rangle\notag\\&=\frac{1}{2}(\sin\theta_\bk\cos\theta_\bk \mathrm{e}^{i\chi}+\sin\theta_\bk\cos\theta_\bk \mathrm{e}^{-i\chi})\langle g,\tilde{g}\vert \tilde{\beta}_{-\mathbf{k}}\tilde{\beta}_{-\mathbf{k}}^\dagger\vert g,\tilde{g}\rangle\notag\\&=\sin\theta_\bk\cos\theta_\bk \cos{\chi}.
\end{align}
Similarly, the other terms are evaluated as follows 
\begin{eqnarray}
\langle S_{\beta \mathbf{k}}^y\rangle=\sin\theta_\bk\cos\theta_\bk \sin{\chi},\quad\langle S^{x2}_{\beta \mathbf{k}}\rangle=\langle S^{y2}_{\beta \mathbf{k}}\rangle=\frac{1}{4},\quad
\langle S_{\beta \mathbf{k}}^z\rangle=\frac{1}{2}(2\sin^2\theta_{\mathbf{k}}-1).
\end{eqnarray}
The left-hand-side of Eq.~(\ref{ct242_1}) is
\begin{align}
(\langle S^{x2}_{\beta \mathbf{k}}\rangle-\langle S_{\beta \mathbf{k}}^x\rangle^2)(\langle S^{y2}_{\beta \mathbf{k}}\rangle-\langle S_{\beta \mathbf{k}}^y\rangle^2)=\frac{1}{16}-\frac{1}{4}\sin^2\theta_{\mathbf{k}}\cos^2\theta_{\mathbf{k}}+\sin^4\theta_{\mathbf{k}}\cos^4\theta_{\mathbf{k}}\sin^2\chi\cos^2\chi.\notag\\
\end{align}
The right-hand-side of Eq.~(\ref{ct242_1}), after some algebra, also gives the same expression.
Therefore, the Robertson-Schrodinger inequality is saturated by the BCS thermal vacuum.
A similar derivation for the $\alpha$-quanta shows the identity also holds. Hence, the BCS thermal vacuum is indeed a generalized squeezed state.



\end{document}